\documentclass[graybox]{svmult}

\usepackage{mathptmx}       
\usepackage{helvet}         
\usepackage{courier}        
\usepackage{type1cm}        
\usepackage{makeidx}         
\usepackage{graphicx}        
\usepackage{multicol}        
\usepackage[bottom]{footmisc}

\makeindex             

\begin{document}

\title*{From Micro- to Macro-scales in the Heliosphere and Magnetospheres}
\author{Dastgeer Shaikh$^1$, I. S. Veselovsky$^{2,3}$, Q. M. Lu$^4$, G.P. Zank$^1$}
\authorrunning{Shaikh et al}
\institute{$^1$Center for Space Plasma and Aeronomic Research and Physics Department, University of Alabama,  Huntsville, AL 35899, USA \email{{\tt dastgeer.shaikh@uah.edu, garyp.zank@gmail.com}}
\and $^2$Skobeltsyn Institute of Nuclear Physics, Moscow State University, Moscow, 119992, Russia
\and $^3$Space Research Institute (IKI), Russian Academy of Sciences, Moscow, 117997, Russia
\and $^4$CAS Key Laboratory of Basic Plasma Physics, School of Earth and Space Sciences, University of Science and Technology of China, Hefei, Anhui, 230026, China}
%
%
\maketitle

\abstract{From a broader perspective, the heliosphere and planetary
  magnetospheres provide a test bed to explore the plasma physics of
  the Universe. In particular, the underlying nonlinear coupling of
  different spatial and temporal scales plays a key role in
  determining the structure and dynamics of space plasmas and
  electromagnetic fields. Plasmas and fields exhibit both laminar and
  turbulent properties, corresponding to either well organized or
  disordered states, and the development of quantitative theoretical
  and analytical descriptions from physics based first principles is a
  profound challenge. Limited observations and complications
  introduced by geometry and physical parameters conspire to
  complicate the problem. Dimensionless scaling analysis and
  statistical methods are universally applied common approaches that
  allow for the application of related ideas to multiple physical
  problems. We discuss several examples of the interplay between the
  scales in a variety of space plasma environments, as exemplified in
  the presentations of the session \textit{From Micro- to Macro-scales
    in the Heliosphere and Magnetospheres.}}

\section{Turbulent spectra in the solar wind and interstellar medium}
\label{sec:1}
The solar wind and interstellar medium is predominantly in a turbulent
state (Marsch E. \& C.-Y. Tu 1995; Goldstein et al 1995; Bruno \&
Carbone 2005) in which low frequency fluctuations are described
typically by a magnetohydrodynamics (MHD) description of
plasma. Nonlinear interactions amongst these fluctuations lead to a
migration of energy in the inertial range that is characterized
typically by a Kolmogorov-like 5/3 spectrum (Frisch 1995). The 5/3
power spectrum is observed frequently, both in the interstellar medium
(ISM) and solar wind (SW). The ubiquity of the turbulence spectrum on
a variety of length scales, leading to a Kolmogorov-like 5/3 law, is
one of the long standing puzzles of classical statistical theories of
turbulence, the origin and nature of which remains a topic of
considerable debate. Owing to its complexity, magnetized plasma
turbulence in general is not only lacking substantially in theoretical
developments because of its analytically intractable nature, but it
also poses computationally a challenging task of resolving multiple
scale flows and fluctuations that are best described
statistically. The fields of plasma and hydrodynamic turbulence have
grown tremendously with the advent of high speed supercomputing and
efficient numerical algorithms. It is not possible to cover all
aspects of the field in this article, and so we concentrate mainly on
the physical processes that lead to the 5/3 spectra in both ISM and SW
plasmas. Understanding energy cascade processes is important
particularly from the point of view of nonlinear interactions across
disparate scales, turbulence transport, wave propagation, heating
processes in the solar wind, structure formation, cosmic ray
scattering, and particle acceleration throughout the heliosphere.

%
\begin{figure}[b]
\sidecaption
\includegraphics[scale=.65]{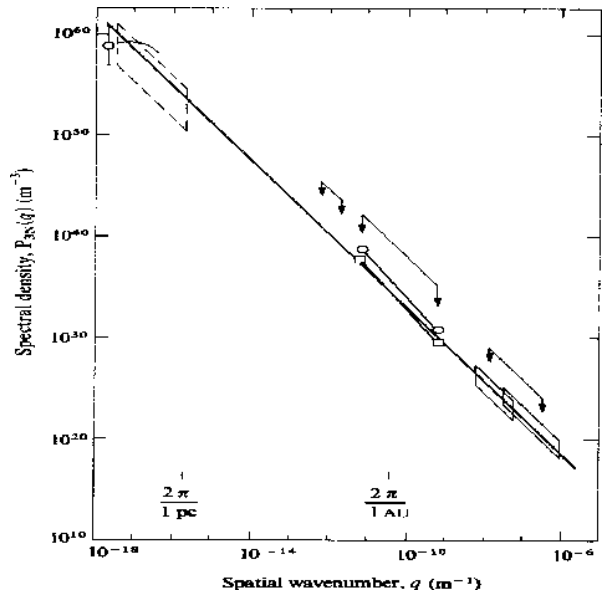}
\caption{ISM turbulence spectrum exhibiting a 5/3 power law (Armstrong et al 1981).}
\label{fig:1}       
\end{figure}

\subsection{Turbulence spectra in the interstellar medium}
It is a curious observation (Fig 1) that electron density fluctuations
in the interstellar medium (ISM) exhibit an omnidirectional
Kolmogorov- like (Kolmogorov 1941) power spectrum k−5/3 (or −11/3
spectra index in three dimensions) over a 4 to 6 decade range
(Armstrong, Cordes \& Rickett 1981; Higdon 1984, 1986; Armstrong et
al. 1990). The observed turbulence spectrum extends over an
extraordinary range of scales i.e. from an outer scale of a few
parsecs to scales of few AUs or less.  Interstellar scintillation,
describing fluctuations in the amplitude and phase of radio waves
caused by scattering in the interstellar medium, exhibit the power
spectrum of the interstellar electron density that follows a 5/3 index
(Armstrong, Rickett \& Spangler 1995). The origin and nature of this
big power law is described in an extensive review by Elmegreen \&
Scalo (2004). Chepurnov \& Lazarian (2010) used the data of the
Wisconsin H$\alpha$ Mapper (WHAM) and determined that the amplitudes
and spectra of density fluctuations can be matched to the data
obtained for interstellar scintillations and scattering that follow a
Kolmogorov-like spectrum spanning from 106 to 1017 m scales.  Angular
broadening measurements also reveal, more precisely, a Kolmogorov-like
power spectrum for the density fluctuations in the interstellar medium
with a spectral exponent slightly steeper than -5/3 (Mutel et al,
1998; Spangler, 1999). Regardless of the exact spectral index, the
density irregularities exhibit a power-law spectrum that is
essentially characteristic of a fully developed isotropic and
statistically homogeneous incompressible fluid turbulence, described
by Kolmogorov (1941) for hydrodynamic and Kraichnan (1965) for
magnetohydrodynamic fluids. Turbulence, manifested by interstellar
plasma fluid motions, therefore plays a major role in the evolution of
the ISM plasma density, velocity, magnetic fields, and the
pressure. Radio wave scintillation data indicates that the rms
fluctuations in the ISM and interplanetary medium density, of possibly
turbulent origin and exhibiting Kolmogorov-like behavior, are only
about 10
2001). This suggests that ISP density fluctuations are only weakly
compressible. Despite the weak compression in the ISP density
fluctuations, they nevertheless admit a Kolmogorov-like power law, an
ambiguity that is not yet completely resolved by any fluid/kinetic
theory or computer simulations. That the Kolmogorov-like turbulent
spectrum stems from purely incompressible fluid theories (Kolmogorov,
1941; Kraichnan, 1965) of hydrodynamics and magnetohydrodynamics
offers the simplest possible turbulence description in an isotropic
and statistically homogeneous fluid. However, since the observed
electron density fluctuations in the ISM possess a weak degree of
compression, the direct application of such simplistic turbulence
models to understanding the ISM density spectrum is not entirely
obvious. Moreover, the ISM is not a purely incompressible medium and
can possess many instabilities because of gradients in the fluid
velocity, density, magnetic field etc. where incompressibility,
inhomogeneity and even isotropy are certainly not good
assumptions. This calls for a fully self-consistent description of ISM
fluid, one that couples incompressible modes with weakly compressible
modes and deals with the strong nonlinear interactions amongst the ISM
density, temperature, velocity and the magnetic field. Note that the
coupling of different modes is an intrinsic property of MHD
perturbations of finite amplitude.

\subsection{Solar wind turbulence spectra}
Solar wind plasma, on the other hand, occurs on much smaller scales,
i.e. few thousands of kilometers, compared to the ISM scales. A wealth
of data from in-situ observations is available from numerous
spacecraft and reveals the nonlinear turbulent character of the
magnetized solar wind plasma fluid. It is evident from these
observations that the solar wind plasma yields a multitude of spatial
and temporal length-scales associated with an admixture of waves,
fluctuations, structures and nonlinear turbulent interactions. In-situ
measurements (Matthaeus \& Brown 1988, Goldstein et al 1994, 1995,
Ghosh et al 1996) indicate that solar wind fluctuations, extend over
several orders of magnitude in frequency and wavenumber. The
fluctuations can be described by a power spectral density (PSD)
spectrum that can be divided into three distinct regions (Goldstein et
al 1995, Leamon et al 1999) depending on the frequency and wavenumber.
This is shown in the schematic of Fig 2. The first region corresponds
to a flatter spectrum, associated with lower frequencies consistent
with a $k^{-1}$ (where $k$ is wavenumber) power law. A second
identifiable region follows and extends to the ion/proton
gyrofrequency, with a spectral slope that has an index ranging from
-3/2 to -5/3. This region is identified with fully developed
turbulence, and is generally described on the basis of the
incompressible magnetohydrodynamic (MHD) equations. The turbulent
interactions in this regime are thought to be governed entirely by
Alfvenic cascades. Spacecraft observations (Leamon et al 1999, Bale et
al. 2005, Alexandrova et al 2007, 2008, Sahraoui et al 2007, 2009)
further reveal that at length scales beyond the MHD regime,
i.e. length scales less than ion gyro radius $k\rho_i<1$ and temporal
scales greater than the ion cyclotron frequency $\omega>\omega_{ci}=e
B_0 / m_ec$ (where $k, \rho_{ci}, e, B_0, m_e, c$ are respectively
characteristic mode, ion gyroradius, ion cyclotron frequency,
electronic charge, mean magnetic field, mass of electron, and speed of
light), the spectrum exhibits a spectral break, and the spectral index
of the solar wind turbulent fluctuations varies between -2 and -5
(Smith et al 2006, Goldstein et al 1994, Leamon et al 1999, Bale et
al. 2005, Shaikh \& Shukla 2008, 2009, Sahraoui et al 2009). Higher
time resolution observations find that at the spectral break, Alfvenic
MHD cascades (Smith et al 2006, Goldstein et al 1994, Leamon et al
1999) close. The characteristic modes in this region appear to evolve
typically on timescales associated with dispersive kinetic Alfvenic
fluctuations.

The onset of the second or the kinetic Alfv\'en inertial range is not
understood. Some suggestions have however been made. The spectral
break may result from energy transfer processes associated with
possibly kinetic Alfven waves (KAWs) (Hasegawa 1976), electromagnetic
ion- cyclotron-Alfven (EMICA) waves (e.g., Gary, 2008) or by
fluctuations described by a Hall MHD (HMHD) plasma model (Alexandrova
et al 2007, 2008; Shaikh \& Shukla 2008, 2008a).  Stawicki et al
(2001) argue that Alfven fluctuations are suppressed by proton
cyclotron damping at intermediate wavenumbers so the observed power
spectra are likely to comprise weakly damped dispersive magnetosonic
and/or whistler waves. Beinroth \& Neubauer (1981) and Denskat \&
Neubauer (1982) have reported the presence of whistler waves based on
Helios 1 and 2 observations in this high frequency regime. A
comprehensive data analysis by Goldstein et al.  (1994), based on
correlations of sign of magnetic helicity with direction of magnetic
field, indicates the possible existence of multiscale waves
(Alfv\'enic, whistlers and cyclotron waves) with a single polarization
in the dissipation regime. Counter-intuitively, in the
$\omega<\omega_{ci}$ regime, or Alfvenic regime, Howes et al. (2008)
noted the possibility that highly obliquely propagating KAWs are
present (with $\omega>\omega_{ci}$) making it questionable that
damping of ion cyclotron waves is responsible for the spectral
breakpoint.

\begin{figure}[ht]
\sidecaption
\includegraphics[scale=.60]{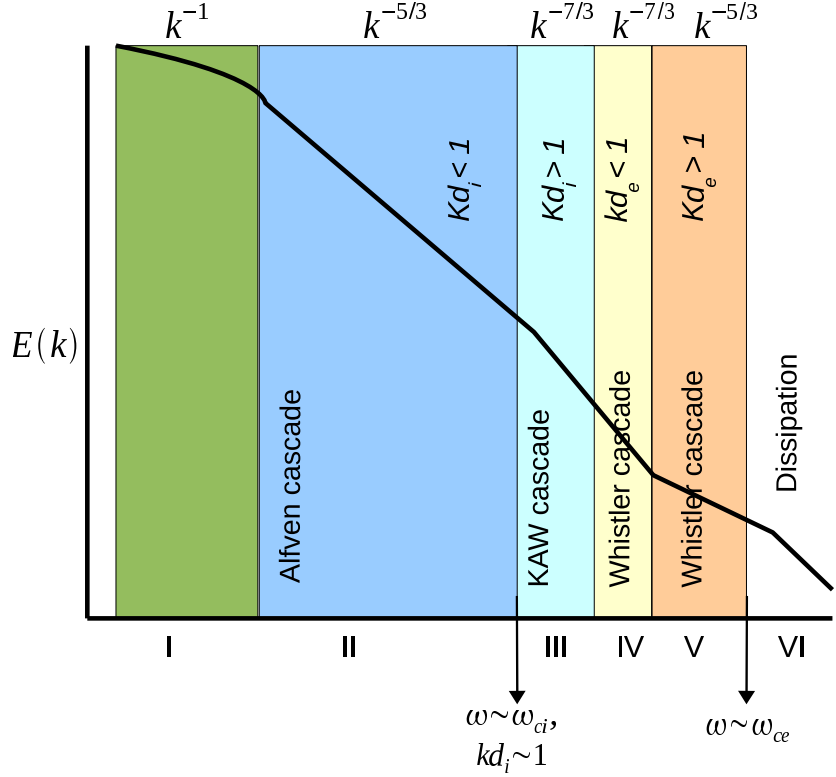}
\caption{ Schematic of the power spectral density (PSD) composite
  spectrum in the solar wind turbulent plasma as a function of
  frequency (wavenumber). Several distinct regions are identified with
  what is thought to be the dominant energy transfer mechanism for
  that particular region. The nonlinear processes associated with the
  transition from region II (MHD regime) to region III (kinetic or
  Hall MHD regime) are not yet fully understood. The power spectra in
  region III vary from $k^{-2}$ to $k^{-4}$ . The boundary of region
  III and IV identifies where electron and ion motions are
  decoupled. Regions IV and V are identified as whistler cascade
  regimes. The outerscale of MHD turbulence corresponds to the smaller
  k mode in region II which can possibly extend over a few parsecs in
  the context of ISM (Armstrong et al 1981).}
\label{fig:2} 
\end{figure}

Fluid (Shaikh \& Zank 2010) and kinetic (Howes et al. 2008)
simulations, in qualitative agreement with spacecraft data described
as above, have been able to obtain the spectral break point near the
characteristic turbulent length scales that are comparable with the
ion inertial length scale ($d_i$). These simulations find
Kolmogorov-like $k^{-5/3}$ spectra for length scales larger than ion
inertial length scales, where MHD is typically a valid description. By
contrast, smaller (than di ) scales were shown to follow a steeper
spectrum that is close to $k^{-7/3}$ (Howes et al. 2008, Shaikh \&
Shukla 2009). Spacecraft data and simulations thus reveal that
migration of turbulent energy proceeds essentially through different
regions in k-space, i.e. $k^{-1}, k^{-5/3}$ and $k^{-7/3}$. Of course,
the turbulent cascade does not entirely terminate immediately beyond
the $k^{-7/3}$ regime.  Fluid and kinetic simulations (Biskamp 1996,
Galtier 2006, Galtier \& Buchlin 2007, Cho \& Lazarian 2003, Shaikh \&
Zank 2005, Shaikh 2009, Gary et al. 2008, Saito et al. 2008, Howes et
al. 2008) show that spectral transfer of energy extends even beyond
the $k^{-7/3}$ regime and is governed predominantly by small scale,
high frequency, whistler turbulence. The latter also exhibits a power
law.

\subsection{Extended composite spectra of the solar wind plasma}
Theory and simulations indicate that turbulent fluctuations in the
high frequency and $k\rho_i >1$ regime correspond to electron motions
that are decoupled from the ion motions (Kingsep et al 1990, Biskamp
et al 1996, Shaikh et al 2000a, Shaikh et al 2000b, Shaikh \& Zank
2003, Cho \& Lazarian 2004, Saito et al 2008, Gary et al
2008). Correspondingly, ions are essentially unmagnetized and can be
treated as an immobile neutralizing background fluid. This regime
corresponds to the whistler wave band of the spectrum and comprises
characteristic scales that are smaller than those that describe MHD,
KAW or Hall MHD processes. An extended composite schematic describing
the whistler mode spectra is also shown in Fig 2. Specifically,
regions IV and V in Fig. 2 identify characteristic modes that are
relevant for the description of whistler wave turbulence (Biskamp et
al 1996, Shaikh \& Zank 2005, Shaikh 2009). The boundary of regions
III and IV represents a wavenumber band in spectral space that
corresponds to the decoupling of electron and ion motions. Wavenumbers
above this boundary characterize the onset of whistler turbulence. The
spectral cascades associated with whistler turbulence are described
extensively by Biskamp et al (1996), Shaikh et al (2000a), Shaikh et
al (2000b), Shaikh \& Zank (2003, 2005), Shaikh (2009a,b). Cho \&
Lazarian 2004 describe scale dependent anisotropy that is mediated by
whistler waves in the context of electron MHD plasma. Gary et al.
(2008) and Saito et al. (2008) have reported two-dimensional
electromagnetic particle-in-cell simulations of an electron MHD model
to demonstrate the forward cascade of whistler turbulence. Their work
shows that the magnetic spectra of the cascading fluctuations become
more anisotropic with increasing fluctuation energy. Interestingly,
whistler turbulence associated with longer wavelengths in region IV
exhibits a power spectrum $k^{-7/3}$ that is similar to the short
wavelength spectrum of kinetic Alfven waves (KAW), as shown in region
III of Fig 2. The underlying physical processes responsible for the
spectrum differ significantly for KAW and whistler waves.

The Hall MHD description of magnetized plasma is valid up to region
III where characteristic turbulent scales are smaller than ion
inertial length scales ($kd_i > 1$). Beyond this location, high
frequency motion of plasma is governed predominantly by electron
motions, and ions form a stationary neutralizing
background. Consequently, the ion motions decouple significantly from
electron motion. These aspects of the spectra, depicted by regions IV
\& V in Fig 2, can be described adequately by whistler wave model. The
Hall MHD models are therefore not applicable in regions IV, V and
beyond. Neither can they describe kinetic physics associated with the
dissipative regime. Since the high frequency regime (i.e. regions IV
\& V) is dominated by electron motions, there exists an intrinsic
length scale corresponding to the electron inertial length scale $d_e=
c /\omega_{p_e}$ (where $\omega_{p_e}$ is the electron plasma
frequency). The characteristic turbulent length scales in regions IV
\& V are comparable with de and therefore describe scales larger
(i.e. $kd_e < 1$ in region IV) and smaller (i.e. $kd_e > 1$ in region
V) than the electron inertial scale. While whistler wave models can
describe nonlinear processes associated with length scales as small as
the electron inertial length scale, they fail to describe finite
electron Larmor radius effects for which a fully kinetic description
of plasma must be used.

Beside those issues described above, we do not understand what leads
to the decoupling of ion and electron motions near the boundary of
region III and IV for example. Although the turbulent spectra are
described by similar spectral indices, the nonlinear processes are
fundamentally different in region III and IV.

\subsection{A nearly incompressible description of the SW and ISM spectra- the 5/3 MHD regime}
Earlier fluid models, describing the turbulent motion of a
compressible ISM fluid, have been based mostly on isothermal and
adiabatic assumptions, due largely to their tractability in terms of
mathematical and numerical analysis. Unfortunately, such models cannot
describe the complex nonlinear dynamical interactions amongst ISM
fluctuations self-consistently. For instance, density fluctuations, in
the context of related solar wind work, were thought to have
originated from nonlinear Alfven modes (Spangler, 1987). A simple
direct relationship of density variations with Alfvenic fluctuations
is not entirely obvious as the latter are not fully self-consistent
and are incompressible by nature thereby ignoring effects due to
magnetoacoustic perturbations for example. On the other hand, fully
compressible nonlinear MHD solutions, for both high- and low-cases,
show that Alfven and slow modes exhibit a $k^{-5/3}$ spectrum, while
fast modes follow a $k^{-3/2}$ spectrum (Cho \& Lazarian 2002, 2003). 
The formation of density power spectrum in the simulations of
isothermal MHD turbulence was studied in Cho \& Lazarian (2003),
Beresnyak, Lazarian \& Cho (2005), Kowal, Lazarian \& Beresnyak
(2008). In particular, in Beresnyak et al. (2005), the logarithm of
density was shown to follow the Goldreich-Sridhar scaling in terms of
both density and anisotropy. This is an important finding that sheds
the light onto the nature of the density fluctuations.

One of the most debated issues in the context of solar wind turbulence
is the non-equipartition between the kinetic and magnetic part of the
energies that leads to a discrepancy between the two spectra.  The
kinetic as well as magnetic energy spectra for fast or slow modes
nevertheless do not relate to a Kolmogorov-like density spectrum. The
latter modes have been suggested as candidates for generating density
fluctuations (Lithwick \& Goldreich , 2001) in the interstellar
medium. Alternate explanations are that density structures
(anisotropic) in the ISM emerge from pressure-balance stationary modes
of MHD (also called Pressure Balance Structures, PBS) (Higdon, 1986),
or from inhomogenities in the large-scale magnetic field via the
four-field model of Bhattacharjee et al. (1998). These descriptions
are inadequate for a general class of ISM problems. The PBSs form a
special class of MHD solutions and are valid only under certain
situations when the magnetic and the pressure fluctuations exert equal
forces in the stationary state. These structures, limited in their
scope to the general ISM conditions, nevertheless do not offer an
entirely self-consistent explanation to the observed density
spectrum. Similarly, an inertial range turbulent cascade associated
with the low turbulent Mach number four field MHD model is not yet
known. Moreover this isothermal inhomogeneous fluid model is valid
only for a class of MHD solutions and yields a linear Mach number
($M$) scaling, $O(M)$, amongst the various fluctuations (Bhattacharjee
et al., 1998).

One of the earlier attempts to understand the ISM density
fluctuations, and relate it to an incompressible fluid turbulence
model dates back Montgomery et al. (1987) who used an assumed equation
of state to relate ISM density fluctuations to incompressible
MHD. This approach, called a pseudosound approximation, assumes that
density fluctuations are proportional to the pressure fluctuations
through the square of sound speed. The density perturbations in their
model are therefore ``slaved'' to the incompressible magnetic field
and the velocity fluctuations. This hypothesis was further contrasted
by Bayly et al. (1992) on the basis of their 2D compressible
hydrodynamic simulations by demonstrating that a spectrum for density
fluctuations can arise purely as a result of abandoning a barotropic
equation of state, even in the absence of a magnetic field. The
pseudosound fluid description of compressibility, justifying the
Montgomery et al. approach to the density-pressure relationship, was
further extended by Matthaeus and Brown (1988) in the context of a
compressible magnetofluid (MHD) plasma with a polytropic equation of
state in the limit of a low plasma acoustic Mach number (Matthaeus and
Brown, 1988). The theory, originally describing the generation of
acoustic density fluctuations by incompressible hydrodynamics
(Lighthill, 1952), is based on a generalization of Klainerman and
Majda's work (Klainerman and Majda, 1981, 1982; Majda, 1984) and
accounts for fluctuations associated with a low turbulent Mach number
fluid, unlike purely incompressible MHD. Such a nontrivial departure
from the incompressible state is termed ``nearly incompressible.'' The
primary motivation behind NI fluid theory was to develop an
understanding and explanation of the interstellar scintillation
observations of weakly compressible ISM density fluctuations that
exhibit a Kolmogorov-like power law. The NI theory is, essentially, an
expansion of the compressible fluid or MHD equations in terms of weak
fluctuations about a background of strong incompressible
fluctuations. The expansion parameter is the turbulent Mach
number. The leading order expansion satisfies the background
incompressible hydrodynamic or magnetohydrodynamic equations (and
therefore fully nonlinear) derived on the basis of Kreiss principle
(Kreiss, 1982), while the higher order yields a high frequency weakly
compressible set of nonlinear fluid equations that describe low
turbulent Mach number compressive HD as well as MHD effects. Zank and
Mathaeus derived the unified self-consistent theory of nearly
incompressible fluid dynamics for non-magnetized hydrodynamics as well
as magnetofluids, with the inclusion of the thermal conduction and
energy effects, thereby identifying different and distinct routes to
incompressibility (Zank \& Matthaeus, 1991, 1993). In the NI theory,
the weakly perturbed compressive fluctuations (denoted by subscript 1)
are expanded about the incompressible modes (denoted by superscript1)
for velocity and pressure variables as $U=U^\infty + \varepsilon U_1$,
$p=p_0+\varepsilon^2 (p^\infty + p_1)$ respectively. Here
$\varepsilon$ is a small parameter associated with the turbulent fluid
Mach number $M_s$ through the relation $C_s^2 = \gamma p/\rho$, $M_s=
U_0 / C_s$ and $\gamma$ is the ratio of the specific heats, $U_0$ is
the characteristic speed of the turbulent fluid, and Cs is the
acoustic speed associated with sound waves. Due to a lack of
uniqueness in the representation of the fluid density and temperature
fields, either of the choices $T=T_0 + \varepsilon T_1$ or $T=T_0 +
\varepsilon^2 T_1$ is consistent.  The first choice corresponds to a
state where temperature fluctuations dominate both the incompressible
and compressible pressure and is referred to as the heat fluctuation
dominated (HFD) regime. On the other hand the second choice in which
all the variables are of similar order is described as the heat
fluctuation modified (HFM) regime. Since the thermal fluctuations in
HFD regime appear at an order $O(\varepsilon)$ as compared with the
pressure $O(\varepsilon^2)$, they dominate the NI ordering. By
contrast, the thermal fluctuations have the 2 same ordering with
respect to the other fluctuations (density, pressure etc) in a HFM
regime. The NI theory introduces a further fundamentally different
explanation for the observed Kolmogorov-type density spectrum in that
the ISM density fluctuations can be a consequence of passive scalar
convection due to background incompressible fluctuations as well as a
generalized pseudo-sound theory. The theory further predicts various
correlations between the density, temperature and the acoustic as well
as convective pressure fluctuations (Zank \& Matthaeus, 1991, 1993,
Shaikh \& Zank 2005, 2006, 2007).

\begin{figure}[ht]
\sidecaption
\includegraphics[scale=.55]{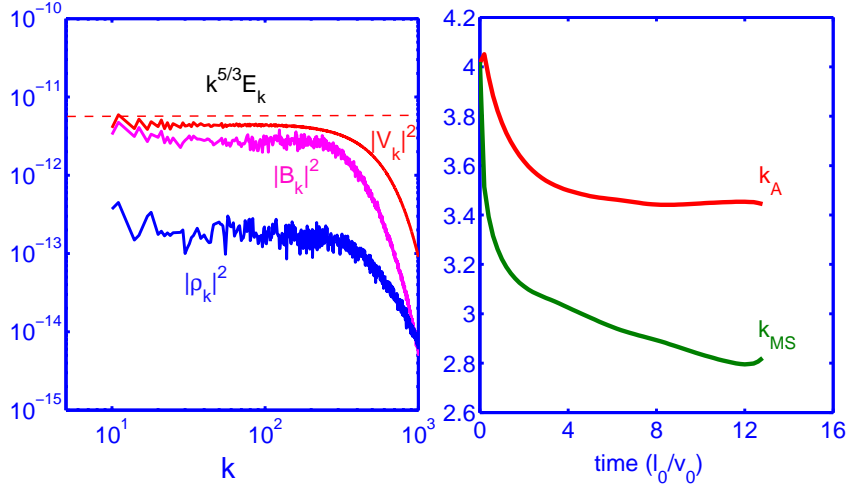}
\caption{(Left) Velocity fluctuations are dominated by shear Alf'enic
  motion and thus exhibit a Kolmogorov-like $k^{-5/3}$ spectrum. The
  middle curve shows the magnetic field spectrum. Density fluctuations
  are passively convected by the nearly incompressible shear Alfvenic
  motion and follow a similar spectrum in the inertial range. The
  numerical resolution in 3D is $512^3$ . (Right) The evolution of
  Alfvenic (kA) and fast/slow magnetosonic (kMS) modes demonstrates
  that the spectral cascades are dominated by Alfvenic modes.  }
\label{fig:3} 
\end{figure}

The validity and nonlinear aspects of the NI model, within the context
of the interstellar medium, has recently been explored by Shaikh \&
Zank (2003, 2005, 2006, 2007, 2010). The theory of nearly
incompressible (NI) fluids, developed by Matthaeus, Zank and Brown,
based on a perturbative expansion technique is a rigorous theoretical
attempt to understand the origin of weakly compressible density
fluctuations in the interstellar medium, and one that provides
formally a complete fluid description of ISM turbulence with the
inclusion of thermal fluctuations and the full energy equation
self-consistently, unlike the previous models described above (Zank \&
Matthaeus, 1990, 1991, 1993; Matthaeus and Brown, 1988). Owing to its
broad perspective and wide range of applicability for interstellar
medium problems, we use here a nearly incompressible description of
fluids to investigate interstellar turbulence with a view to
explaining the observed Kolmogorov-like ISM density spectrum. A
central tenant of the homogeneous NI theory is that the density
fluctuations are of higher order, of higher frequency and possess
smaller length-scales than their incompressible counterparts to which
they are coupled through passive convection and the low frequency
generation of sound. Most recently, Hunana \& Zank 2006, 2010 have
extended the NI hydrodynamic and MHD theory to inhomogeneous flows,
finding that the density fluctuations can also be of order Mach
number, in agreement with a slightly different approach advocated by
Bhattacharjee et al., 1998. The NI fluid models, unlike fully
incompressible or compressible fluid descriptions, allow us to address
weakly compressible effects directly in a quasi-neutral ISM
fluid. Furthermore, NI theory has enjoyed notable success in
describing fluctuations and turbulence in the supersonic solar wind.
The NI model has recently been solved numerically and compared to
observations in an effort to understand the Kolmogorov-like density
spectrum in the ISM (Shaikh \& Zank 2004, 2005, 2006, 2007, 2008,
2009, 2010). One of our results, shown in Fig 3, describes the
evolution of density fluctuations from a fully compressible initial
state. We find from our three-dimensional (decaying turbulence)
simulations that a $k^{-5/3}$ density fluctuation spectrum emerges in
fully developed compressible MHD turbulence from non-linear mode
coupling interactions that lead to the migration of spectral energy in
the transverse (i.e. $k \perp U$) Alfvenic fluctuations, while the
longitudinal ``compressional modes'' corresponding to $k \parallel U$
fluctuations make an insignificant contribution to the spectral
transfer of inertial range turbulent energy. The explanation, in part,
stems from the evolutionary characteristics of the MHD plasma that
governs the evolution of the non-solenoidal velocity field in the
momentum field. It is the non-solenoidal component of plasma motions
that describes the high-frequency contribution corresponding to the
acoustic time-scales in the modified pseudo-sound relationship
(Montgomery et al. 1987; Matthaeus et al.  1998; Zank \& Matthaeus
1990, 1993). What is notable in the work of Shaikh \& Zank is that
they find a self-consistent evolution of a Kolmogorov-like density
fluctuation spectrum in MHD turbulence that results primarily from
turbulent damping of non-solenoidal modes that constitute fast and
slow propagating magnetoacoustic compressional perturbations. These
are essentially a higher frequency (compared with the Alfvenic waves)
component that evolve on acoustic timescales and can lead to a
``pseudo-sound relationship'' as identified in the nearly
incompressible theory (Zank \& Matthaeus 1990, 1993; Bayly et
al. 1992; Matthaeus et al. 1998; Shaikh \& Zank 2004a,b,c, 2006,
2007). The most significant point to emerge from the simulation is the
diminishing of the high-frequency component that is related to the
damping of compressible plasma motion. This further leads to the
dissipation of the small-scale and high- frequency compressive
turbulent modes. Consequently, the MHD plasma relaxes towards a nearly
incompressible state where the density is convected passively by the
velocity field and eventually develops a $k^{-5/3}$ spectrum. This
physical picture suggests that a nearly incompressible state develops
naturally from a compressive MHD magnetoplasma in the solar wind.

Among other work, describing a Kolmogorov-like 5/3 spectrum in the
context of MHD turbulence, are Cho \& Vishniac (2000), Maron \&
Goldreich (2001), Cho, Lazarian \& Vishniac (2002, 2003), Muller \&
Biskamp (2002), Cho \& Lazarian (2002, 2003), Kritsuk et al. (2009).
Our results, describing a Kolmogorov-like 5/3 spectrum in the solar
wind plasma, are thus consistent with these work. It is noted,
however, that Maron \& Goldreich (2001) report a Kraichnan-like 3/2
spectrum in the inertial range for velocity fluctuations. The
controversy of 5/3 or 3/2 is nonetheless beyond the scope of our
review article.

\subsection{Hall MHD model of SW turbulence- Extended spectra}
To describe the extended solar wind spectra in Fig 2, time dependent,
fully compressible three dimensional simulations of Hall MHD plasma in
a triply periodic domain have been developed.  This represents a local
or regional volume of the solar wind plasma or ISP. The turbulent
interactions in region II were described above by a 3D MHD model which
is a subset of Hall MHD model since it does not contain the ${\bf J}
\times {\bf B}$ term in the magnetic field induction equation.  Note
that the dynamics of length-scales associated with region III,
i.e. corresponding to the KAW modes, cannot be described by the usual
MHD models as they do not describe turbulent motions corresponding to
the characteristic frequencies larger than an ion gyro frequency. At 1
AU, ion inertial length scales are smaller than ion gyro radii in the
solar wind (Goldstein 1995).  Plasma effects due to finite Larmor
radii can readily be incorporated in MHD models by introducing Hall
terms to accommodate ion gyro scales up to scales as small as ion
inertial length scales.

The Hall model in the limit of a zero ion-inertia converges to the
usual MHD model, and assumes that the electrons are inertial-less,
while the ions are inertial (Mahajan \& Krishan 2004).  Hence, the
electrons and ions have a differential drift, unlike the one fluid MHD
model for which the electron and ion flow velocities are
identical. The Hall MHD description of magnetized plasma has
previously been employed to investigate wave and turbulence processes
in the context of solar wind plasma. Sahraoui et al. (2007) extended
the ordinary MHD system to include spatial scales down to the ion skin
depth or frequencies comparable to the ion gyrofrequency in an
incompressible limit. They further analyzed the differences in the
incompressible Hall MHD and MHD models within the frame work of linear
modes, their dispersion and polarizations. Galtier (2006) developed a
wave turbulence theory in the context of an incompressible Hall MHD
system to examine the steepening of the magnetic fluctuation power law
spectra in the solar wind plasma. Furthermore, Galtier \& Buchlin
(2007) have developed 3D dispersive Hall magnetohydrodynamics
simulations within the paradigm of a highly turbulent shell model and
demonstrated that the large-scale magnetic fluctuations are
characterized by a $k^{-5/3}$-type spectrum that steepens at scales
smaller than the ion inertial length $d_i$ to $k^{-7/3}$. The observed
spectral break point in the solar wind plasma, shown by the regime III
in Fig 2, has been investigated using 3D simulations of a two fluid
nonlinear Hall MHD plasma model (Shaikh \& Zank, 2009).

In the inertial-less electron limit the electron fluid does not
influence the momentum of solar wind plasma directly except through
the current. Since the electron fluid contributes to the electric
field, plasma currents and the magnetic field are affected by electron
oscillations. The combination of electron dynamics and ion motions
distinguishes the Hall MHD model from its single fluid MHD
counterpart. Thanks to the inclusion of electron dynamics, Hall MHD
can describe solar wind plasma fluctuations that are associated with a
finite ion Larmor radius and thus a characteristic plasma frequency is
$\omega>\omega_{ci}$. Because Hall MHD contains both ion and electron
effects, there is a regime at which the one set of plasma fluctuations
no longer dominates but instead is dominated by the other. This
introduces an intrinsic scale length/timescale (frequency) that
separates ion dominated behavior in the plasma from electron
dominated. It is the Hall term corresponding to the ${\bf J} \times
{\bf B}$ term in Faraday's equation that is primarily responsible for
decoupling electron and ion motion on ion inertial length and ion
cyclotron time scales (and introducing an intrinsic length scale). It
is this feature that makes Hall MHD useful in describing dissipative
solar wind processes when single fluid MHD is not applicable (the MHD
model breaks down at $\omega>\omega_{ci}$). Hall MHD allows us to
study inertial range cascades beyond $\omega>\omega_{ci}$, and can be
extended to study dissipative heating processes where ion cyclotron
waves are damped. However to study MHD processes, once can put $d_i=0$
in region II. The extreme limit of fluid modeling applied to solar
wind processes (even beyond the limit of the Hall MHD regime) is to
use of an electron MHD model in which high frequency electron dynamics
is treated by assuming stationary ions that act to neutralize the
plasma background.

\begin{figure}[ht]
\sidecaption
\includegraphics[scale=.65]{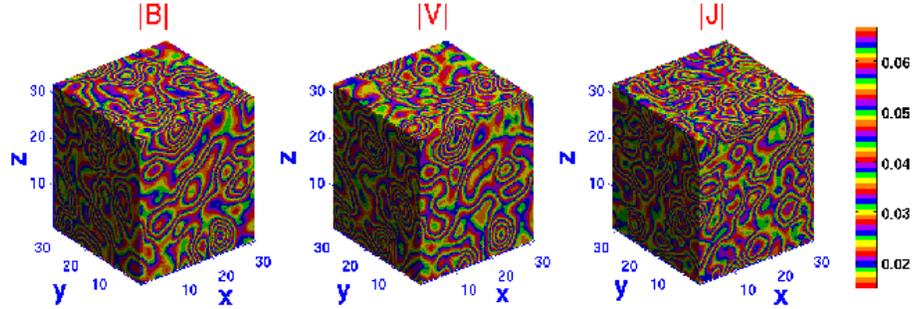}
\caption{ Three dimensional structures of magnetic, velocity and
  current fields in Hall MHD turbulence. Turbulent equipartition
  between velocity and magnetic field leads to almost similar large
  scale structures in the two fields, while current is more
  intermittent. }
\label{fig:4} 
\end{figure}

Turbulence involves nonlinear interactions of modes in all three
spatial directions. Three dimensional computations are numerically
expensive, but, with the advent of high speed vector and parallel
distributed memory clusters, and efficient numerical algorithms such
as those designed for Message Passing Interface (MPI) libraries, it is
now possible to perform magnetofluid turbulence studies at
substantially higher resolutions. Based on MPI libraries, three
dimensional, time dependent, compressible, non-adiabatic, driven and
fully parallelized Hall magnetohydrodynamic (MHD) nonlinear codes have
been developed that run efficiently on both distributed memory
clusters like distributed-memory supercomputers or shared memory
parallel computers. This allows for very high resolution in Fourier
spectral space. Shaikh \& Zank (2010) have developed a 3D periodic
code that is scalable and transportable to different cluster machines,
and extends earlier MHD codes of theirs (Shaikh \& Zank ,2006, 2007,
2008, 2009, 2010).  Their code treats the solar wind plasma
fluctuations as statistically isotropic, locally anisotropic,
homogeneous and random, consistent with ACE spacecraft measurements
(Smith et al 2006).  The numerical algorithm accurately preserve the
ideal rugged invariants of fluid flows, unlike finite difference or
finite volume methods. The conservation of ideal invariants (energy,
enstrophy, magnetic potential, helicity) in inertial range turbulence
is an extremely important feature because these quantities describe
the cascade of energy in the inertial regime, where turbulence is, in
principle, free from large-scale forcing as well as small scale
dissipation.  Damping of plasma fluctuations may nonetheless occur as
a result of intrinsic non-ideal effects such those introduced by the
finite Larmor radius. An example of the plasma velocity, magnetic
field and current is shown in Fig 4.

\begin{figure}[ht]
\sidecaption
\includegraphics[scale=.55]{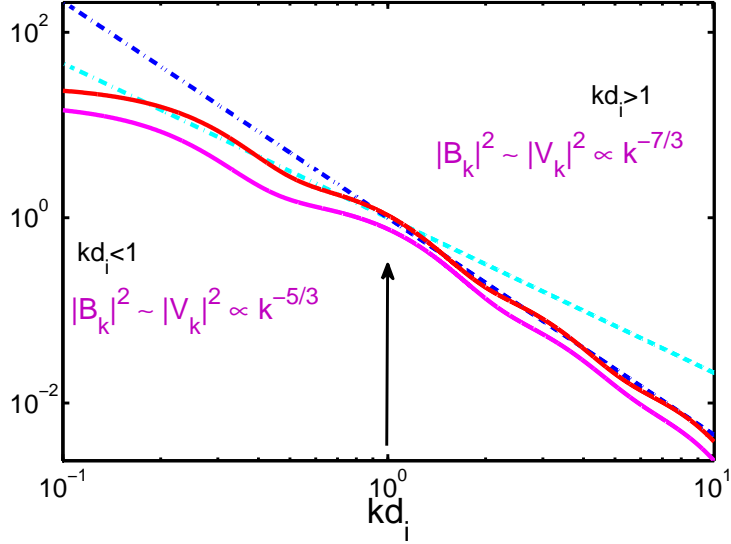}
\caption{ Inertial range turbulent spectra for magnetic and velocity
  field fluctuations. The fluctuations closely follow respectively
  $k^{-5/3}$ and $k^{-7/3}$ scaling in the $kd_i < 1$ and $kd_i > 1$
  KAW regimes. $kd_i = 0.05$ and $1.0$ respectively in the $kd_i < 1$
  and $kd_i > 1$ regimes. The dash-dot straight lines correspond to a
  $k^{-5/3}$ and $k^{-7/3}$ power law.}
\label{fig:5} 
\end{figure}

In the simulations of Shaikh \& Zank, the nonlinear spectral cascade
in the modified KAW regime leads to a secondary inertial range in the
vicinity of $kd_i\simeq 1$ , where the turbulent magnetic and velocity
fluctuations form spectra close to $k^{-7/3}$. This is displayed in
Fig 5, which also shows that for length scales larger than the ion
thermal gyroradius, an MHD inertial range spectrum close to $k^{-5/3}$
is formed. The characteristic turbulent spectrum in the KAW regime is
steeper than that of the MHD inertial range. Identifying the onset of
the secondary inertial range has been the subject of debate because of
the presence of multiple processes in the KAW regime that can mediate
the spectral transfer of energy. These processes include, for
instance, the dispersion and damping of EMICA waves, turbulent
dissipation, etc.

Regimes IV and V, shown in the schematic of Fig 2, requires that we
invoke a whistler model for the plasma. Whistler modes are excited in
the solar wind plasma when the characteristic plasma fluctuations
propagate along a mean or background magnetic field with frequency
$\omega>\omega_{ci}$ and the length scales are $c/\omega_{pi} < \ell
<c/\omega_{pe}$, where $\omega_{pi}, \omega_{pe}$ are the plasma ion
and pi pe electron frequencies respectively. The electron dynamics
plays a critical role in determining the nonlinear interactions while
the ions provide a stationary neutralizing background against fast
moving electrons and behave as scattering centers. Whistler wave
turbulence can be described by an electron magnetohydrodynamics (EMHD)
model for the plasma (Kingsep et al 1990), utilizing a single fluid
description of quasi neutral plasma. The EMHD model has been discussed
in considerable detail in earlier work (Kingsep et al. 1990; Biskamp
et al 1996; Shaikh et al.  2000a; Shaikh et al. 2000B; Shaikh \& Zank
2003; Shaikh \& Zank 2005). In whistler modes, the currents carried by
the electron fluid are important (Shaikh 2000, 2009, 2010). Turbulent
interactions mediated by the coupling of whistler waves and inertial
range fluctuations have been studied in three dimensions based on a
nonlinear 3D whistler wave turbulence code (Shaikh \& Zank, 2010).

Electron whistler fluid fluctuations, in the presence of a constant
background magnetic field, evolve by virtue of nonlinear interactions
in which larger eddies transfer their energy to smaller eddies through
a forward cascade. The Kolmogorov model postulates that the cascade of
spectral energy occurs exclusively between neighboring Fourier modes
(i.e. local interaction) until the energy in the smallest turbulent
eddies is finally dissipated. This leads to a damping of small
scale motions. By contrast, the large-scales and the inertial range
turbulent fluctuations remain unaffected by direct dissipation of the
smaller scales. IN the absence of a mechanism to drive turbulence at
the larger scales in the Shaikh \& Zank 2009 simulations, the
large-scale energy simply migrates towards the smaller scales by
virtue of nonlinear cascades in the inertial range and is dissipated
at the smallest turbulent length-scales. The spectral transfer of
turbulent energy in the neighboring Fourier modes in whistler wave
turbulence follows a Kolmogorov phenomenology (Kolmogorov 1941,
Iroshnkov 1963, Kraichnan 1965) that leads to Kolmogorov-like energy
spectra. Thus, the 3D simulations of whistler wave turbulence in the
$kd_e< 1$ and $kd_e > 1$ regimes exhibits respectively $k^{-7/3}$ and
$k^{-5/3}$ (see Fig 6) spectra. The inertial range turbulent spectra
obtained from 3D simulations are consistent with 2D models (Shaikh \&
Zank 2005). The whistler wave dispersion relation shows that wave
effects dominate at the large scale, i.e. the $kd_e < 1$ regime, and
the inertial range turbulent spectrum exhibits a Kolmogorov-like
$k^{-7/3}$ spectrum. On the other hand, turbulent fluctuations on
smaller scales (i.e., in the $kd_e > 1$ regime) behave like
non-magnetic eddies in a hydrodynamic fluid and yield a $k^{-5/3}$
spectrum.  The wave effect is weak, or negligibly small, in the
latter. Hence the nonlinear cascades are determined essentially by the
hydrodynamic-like interactions. Thus, the observed whistler wave
turbulence spectra in the $kd_e < 1$ and $kd_e > 1$ regimes (Figs 6)
can be understood on the basis of Kolmogorov-like arguments that
describe the inertial range spectral cascades. In the electron
whistler wave regime, the fluid simulations describing a 7/3 spectrum
are also reported by Ng et al. (2003), Cho \& Lazarian (2004, 2010).
Their results are consistent with our simulations described in 
Fig (6a).

\begin{figure}[ht]
\sidecaption
\includegraphics[scale=.28]{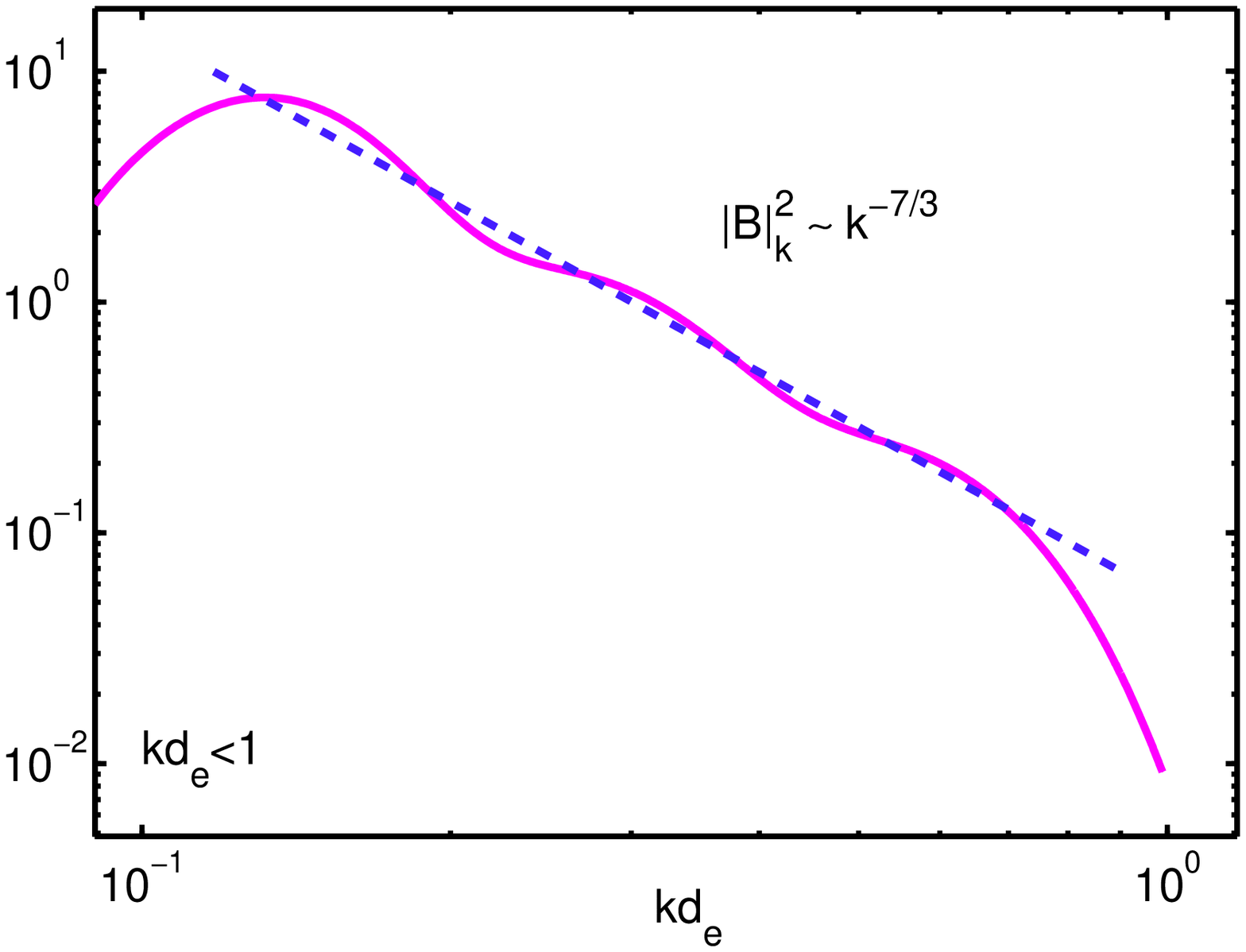}
\includegraphics[scale=.285]{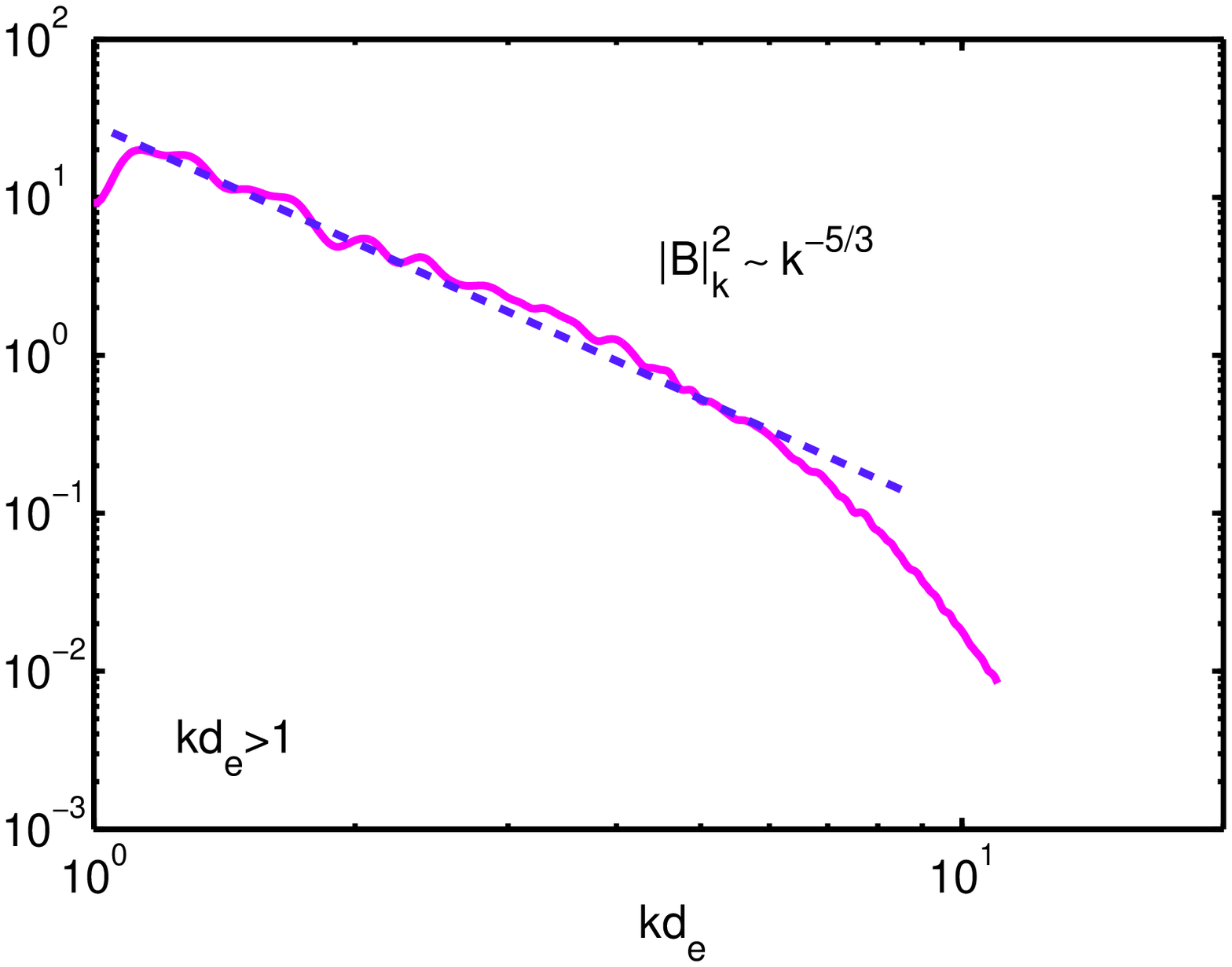}
\caption{ (Left) 3D simulation of whistler wave turbulence in the
  $kd_e < 1$ regime exhibits a Kolmogorov-like inertial range power
  spectrum close to $k^{-7/3}$. (Right) The small scales magnetic
  field fluctuations in the $kd_e > 1$ regime depicts a
  Kolmogorov-like $k^{-5/3}$ spectrum which is a characteristic of
  hydrodynamic fluid.}
\label{fig:6} 
\end{figure}

We note that 7/3 regime of whistler turbulence is different from the
usual 5/3 regime in the MHD turbulence. The 5/3 regime does not
terminate sharply beyond the inertial range MHD fluctuations, but
there is another cascade regime, not describable by the MHD equations,
that deviates significantly from the 5/3 regime and is describable by
whistler mode turbulence.

\section{Perpendicular shocks}
Diffusive shock acceleration (DSA) is considered to be the mechanism
responsible for the acceleration of energetic particles and the
consequent generation of power-law spectra observed at quasi-parallel
shocks (Axford et al., 1977; Bell 1978; Blandford \& Ostriker, 1978).
At a quasi- parallel shock, energetic ions can be scattered by the
self-excited and pre-existing waves and turbulence upstream and
downstream of the shock, so leading to their multiple crossing of the
shock. Because these ions can stream far upstream along the magnetic
field and excite low- frequency plasma waves, the turbulence
responsible for particle scattering ahead of the shock is present. In
this way, energetic particles can be accelerated by DSA to high
energies and form a power-law spectrum (Lee, 1983; Zank et al, 2000).

At a quasi-perpendicular shock, no self-consistent plasma wave
excitation occurs upstream, which therefore limits particle
scattering. Because of this, DSA cannot be used to explain the
observed power-law spectra of energetic particles at
quasi-perpendicular shock waves in the usual way. Lu et al. (2009)
investigated the interaction of Alfven waves with a perpendicular
shock using a two-dimensional hybrid simulation. Alfven waves are
injected from the left boundary, and they have no obvious effects on
the propagation speed of the shock. After the upstream Alfven waves
are transmitted into the downstream, their amplitude is enhanced by
about 10-30 times. Consistent with the fluid theory (McKenzie \&
Westphal, 1969), the transmitted waves can be separated into two
parts: one that propagates along the direction parallel to the
background magnetic field, and the other along the direction
anti-parallel to the background magnetic field. In addition, we also
find obvious ripples in the shock front due to the interaction of the
Alfven waves and the perpendicular shock.  

In a realistic shock, of course, the structure of a
quasi-perpendicular shock is considerably more complicated than
described above, and the meandering magnetic field lines can cross the
shock front more than once. K\'ota (2009) has discussed the efficiency
of ion acceleration at a perpendicular shock using an analytical
approximation and numerical simulations. Energetic ions are generated
at places where the field lines cross into the upstream region and
soon re-cross the shock back to the downstream region. These ions may
be accelerated to very high energies through multiple mirroring at the
stronger downstream field.

Umeda et al. (2009) also discussed the effect of the rippling of
perpendicular shock fonts on electron acceleration in the
shock-rest-frame using a full particle simulation. The cross-scale
coupling between ion-scale mesoscopic shock ripples and an
electron-scale microscopic instability was found to play an important
role in energizing electrons at quasi-perpendicular shocks. At the
shock front, the ions reflected by the shock experience considerable
acceleration upstream at a localized region where the shock-normal
electric field of the rippled structure is polarized upstream. The
current-driven instability is unstable and large-amplitude
electrostatic waves grow upstream. As a result, electrostatic waves
can trap electrons upstream, and then energetic electrons are
generated via a form of surfing acceleration at the leading edge of
the shock transition region.

\begin{figure}[ht]
\sidecaption
\includegraphics[scale=.50]{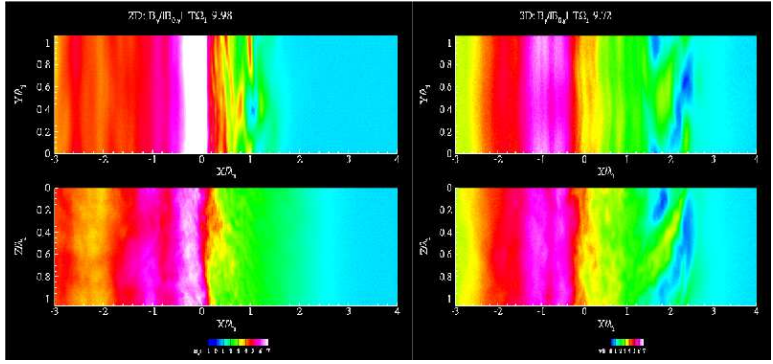}
\caption{ Color contours of the By component (the major component of
  the magnetic field), showing (a) two 2D results (in the XY and XZ
  planes), and (b) a 3D result. The 3D results show that large
  amplitude wave active exists persistently (independent of the
  reformation phase) in the furthest front of the shock. [ Shinohara
    and Fujimoto (2009).]}
\label{fig:7} 
\end{figure}

Shinohara \& Fujimoto (2009) discussed non-stationary behavior of the
shock front, since it has been thought to play an important role for
dissipation mechanism in collision-less shocks. Using JAXA's new
super-computer facility allowed them to perform a three-dimensional
simulation of a quasi-perpendicular shock. The simulation parameters
were selected to simulate specific Cluster observational results. The
full ion to electron mass ratio, $M/m=1840$, was used, and almost (one
ion inertia length)$^2$ square plane perpendicular to the upstream
flow direction was allocated for this simulation. The 3D results of
Shinohara \& Fujimoto (2009) showed that both self-reformation and
whistler emission are present. By comparing their 3D results with 2D
simulations based on the same simulation parameters (Fig. 7), they
confirmed that the 3D result is not simply a superposition of 2D
behavior but instead identified new wave activity in the front of the
shock foot region. Because of the enhanced wave activity, electrons
are much more efficiently heated in the 3D simulations than in 2D
simulations. That the shock structure is changed significantly in
adding a further degree of freedom with the third spatial dimension
emphasizes the importance of fully multi-dimensional studies. The
simulation of Shinohara \& Fujimoto (2009) also identifies the
importance of using the full mass ratio in simulations.

\section{Global magnetospheric modeling and observations}
A systematic evaluation of ground and geostationary magnetic field
predictions generated by a set of global MHD models shows that a
metrics analysis of two different geospace parameters, the
geostationary and ground magnetic field, yields surprising
similarities. However, the parameters reflect rather different
properties of geospace (Pulkkinen et al., 2010). More specifically, by
increasing the spatial resolution and including more realistic inner
magnetospheric physics made the model predictions by the BATS-R-US
model more accurate.  By contrast, the OpenGGCM model had a tendency
to generate larger differences to observations than BATS-R-US in terms
of the prediction efficiency, but the model provided more accurate
representation of the observed spectral characteristics of the ground
and geostationary magnetic field fluctuations. This suggests that both
models capture some of the intrinsic physical elements necessary to
realistic modeling, but the complexity of identifying realistic
boundary conditions and the capturing of the physics between different
plasma regimes in the Earth-magnetospheric interaction means that this
will remain an outstanding problem for years to come.

It is well known that the southward component of the interplanetary
magnetic field (IMF) Bz is the primary heliospheric parameter
responsible for geomagnetic storms. Yakovchouk et al.  (2009)
performed a statistical analysis of the peak values of the IMF $B_z$
component with different combinations of plasma parameters and the
hourly Dst (omniweb.gsfc.nasa.gov) and $Dxt/Dcx$ (Karinen and Mursula,
2005, 2006) geomagnetic indices for all identified perturbations in
1963- 2009 (Fig. 8a). Storms without available interplanetary data
were not included in the database.  Yakovchouk et al. (2009) concluded
that the storms occur more often (twice as often) during the
development phase of the solar cycle than during the rising phase. The
average waiting time between consecutive Dst peaks is 11 days for Dst
< -50 nT and 50 days for $Dst <-100nT$. The average delay time between
Dst and Bz peak values is 4-6 hours. A semiannual variation of the Dst
peak values exists for all levels. Empirical formulae are derived by
Yakovchouk et al. (2009) that relate $Dst/Dcx/Dxt < -50 nT$ and
$B_z/E_y$ values ($E_y=U_x B_z$ - the peak value of electric field,
whereUx is the radial velocity component of the solar wind) based on
their analysis of the observations. The relations that they present
are in a good agreement with the Akasofu relation (Akasofu, 1981), and
are useful for quick estimates and reconstruction of heliospheric and
geomagnetic parameters with accuracy of the order of a few tens
percent (Fig. 8a). A dependence of the area for the development phase
duration of storms with Dst and Bz peaks also exists (Fig.  8b). The
accuracy of reconstruction is less when only fragmentary geomagnetic
data are available.

\begin{figure}[ht]
\sidecaption
\includegraphics[scale=.50]{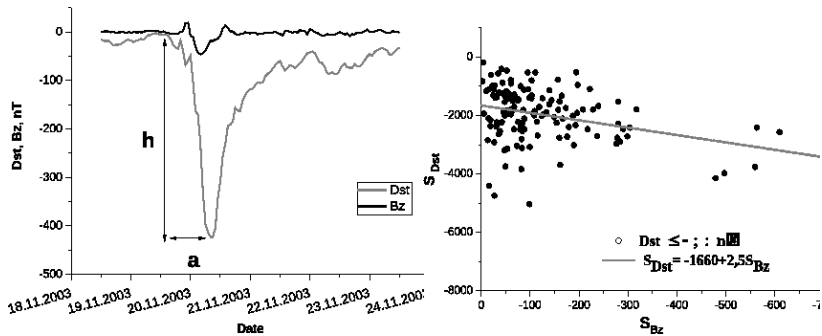}
\caption{ (a) An example of a large geomagnetic storm showing the Bz
  and Dst profiles as a function of time. (b) The dependence of area
  (S=h x a) for the development phase duration of a storm for Dst and
  Bz peaks in 1963-2009 and its approximation by a linear
  fit. (Yakovchouk et al., 2009.)}
\label{fig:8} 
\end{figure}

\section{Distribution functions of protons and interstellar hydrogen in the inner and outer heliosheath}

The Interstellar Boundary EXplorer (IBEX) (McComas et al., 2006,
McComas et al., 2009), launched on 19 October, 2008, is measuring the
energetic neutral atom (ENA) flux from the boundary regions of our
heliosphere. Contemporaneously, Voyager 1 and 2 (V1, V2) are making in
situ measurements of plasma, energetic particles, and magnetic fields
along two trajectories in the heliosheath (Stone et al., 2008,
Richardson et al., 2008, Decker et al., 2008, Burlaga et al., 2008,
Gurnett et al., 2008). The interpretation of the IBEX observations
will depend critically on global simulations of the solar wind-local
interstellar medium (LISM) interaction e.g., Heerikhuisen et al.,
2008, informed by in situ data returned by the Voyager spacecraft.
Underlying the determination of the ENA flux observed at 1 AU is the
form of the proton distribution function in the inner and outer
heliosheath. ENAs are created by charge exchange of interstellar
neutral H and heliosheath (inner and outer) protons or ions. Because
the inner heliosheath is hot, a population of energetic neutral atoms
is created. The flux of ENAs will therefore depend quite sensitively
on the number of particles in the wings of the hot proton population
downstream of the heliospheric termination shock (TS), something
recognized by both Prested et al., 2008 and Heerikhuisen et al., 2008
in their introduction of a $\kappa$-distribution to model the inner
heliosheath proton distribution. In particular, in an important
extension of their earlier work, Heerikhuisen et al., 2008 developed a
fully self-consistent 3D MHD-kinetic neutral hydrogen (H) model
describing the solar wind-LISM interaction (Pogorelov et al., 2006,
Pogorelov et al., 2007, Pogorelov et al., 2008) using a
$\kappa$-distribution to describe the underlying proton distribution
in the inner heliosheath. Previous models assumed a Maxwellian
description for the protons with self-consistent coupling to
interstellar neutral H -- the self-consistent coupling being crucial
in determining the global heliospheric structure (Baranov \& Malama,
1993, Pauls \& Zank, 1995, Zanket al., 1996a, Pogorelov et al., 2006)
(see Zank 1999 ,Zank et al., 2009 and Pogorelov et al., 2009 for
extensive reviews). Prested et al., 2008 by contrast used a test
particle approach to model the neutral H production based on an ideal
MHD model.

The treatment of the heliospheric proton distribution function as a
$\kappa$-distribution yields important differences in both the global
structure of the heliosphere (decreasing the overall extent of the
inner heliosheath between the TS and the heliopause) and the predicted
ENA flux at 1 AU (Heerikhuisen et al., 2008). By assuming a
$\kappa$-distribution with index $\kappa= 1.63$ (this motivated by the
observed spectral index associated with energetic particles downstream
of the heliospheric termination shock, Decker et al., 2005),
Heerikhuisen et al. 2008 find that the ENA flux at 1 AU is
substantially higher than for a corresponding Maxwellian proton
distribution with the same temperature. This is not especially
surprising of course because the $\kappa$-distribution contains many
more particles in the wings of the distribution than the corresponding
Maxwellian, thereby giving higher fluxes of ENAs at higher
energies. Why the heliosheath proton distribution function should be
like a kappa distribution with a spectral index close to 1.63 is
however quite unclear. The answer may well reside in the processing of
the upstream pickup ion distribution by the TS and the subsequent
statistical relaxation of the processed distribution in the
heliosheath (Livadiotis \& McComas, 2009). IBEX will provide
definitive observations of the ENA flux at 1 AU that will allow us to
estimate the proton distribution in the inner heliosheath.

Related to the question of the heliosheath proton distribution are the
plasma and magnetic field observations made by Voyager 2 on the second
crossing of the TS. V2 has a working plasma instrument and the
coverage was sufficient to identify three distinct crossings of the TS
and make in situ measurements of the microstructure. The identified
TS-3 crossing revealed an almost classical perpendicular shock
structure (Burlaga et al., 2008, Richardson et al., 2008).  However,
plasma measurements revealed that the solar wind proton temperature
changed from ~20,000 K upstream to ~ 180,000 K downstream (Richardson
et al., 2008, Richardson, 2009).  Although hot solar wind plasma is
sometimes observed, the average downstream proton plasma temperature
is an order of magnitude smaller than predicted by the MHD
Rankine-Hugoniot conditions, and the global self-consistent models all
yield downstream proton temperatures of $\sim 2 \times 10^6$ K (Zank
et al. 2009). The downstream shock heated solar wind ion temperature
observed by V2 is in fact so low that the downstream flow appears to
remain supersonic (Richardson et al., 2008)!  Furthermore, the
transmitted solar wind proton distribution appears to be essentially a
broadened/heated Maxwellian (with a somewhat flattened peak), and
there is no evidence of reflected solar wind ions being transmitted
downstream (Richardson, 2009). Richardson et al.  2008 and Richardson
2009 concluded that pickup ions (PUIs) experienced preferential
heating at the TS and thus provided both the primary shock dissipation
mechanism and the bulk of the hot plasma downstream of the
TS. Unfortunately, the Voyager spacecraft were not instrumented to
measure PUIs directly. That PUIs provide the TS dissipation and heated
downstream plasma had in fact been predicted by Zank et al. 1996b in
their investigation of the interaction of PUIs and solar wind ions
with the TS. They concluded that ``P[U]Is may therefore provide the
primary dissipation mechanism for a perpendicular TS with solar wind
ions playing very much a secondary role.'' Thus the basic model of
Zank et al. 1996b for the microstructure of the TS appears to be
supported by the V2 observations. However, both the observed solar
wind proton distribution and a shock dissipation mechanism based on
PUIs means that the downstream proton distribution function is a
(possibly complicated) function of the physics of the TS. Zank et al.,
2010 have extended their basic model of the quasi-perpendicular TS,
mediated by PUIs, to derive the complete downstream proton
distribution function in these regions, identifying the partitioning
of energy between solar wind protons and PUIs, and infered potential
implications of these results for the ENA flux observed at 1 AU in
terms of spectra and skymaps. They did not attempt to synthesize a
complete description of the inner heliosheath proton distribution at
this point, preferring instead to elucidate the physics of the
quasi-perpendicular termination shock, and relate that physics to the
production of ENAs. Other regions of the TS, notably the high polar
regime and possibly the heliotail region of the TS, may require the
introduction of distinctly different physical processes for shock
dissipation, and a complete model of the heliosheath proton
distribution will therefore need to account for multiple shock
regimes.  Katushkina \& Izmodenov 2009 have begun to explore different
aspects of this.

The model developed by Zank et al., 2010 describes the basic plasma
kinetic processes and microphysics of the quasi-perpendicular TS in
the presence of an energetic PUI population. They find that the solar
wind protons do not experience reflection at the cross-shock potential
of the TS, and are transmitted directly into the heliosheath. PUIs, by
contrast, can be either transmitted or reflected at the TS, and
provide the primary dissipation mechanism at the shock, and dominate
the downstream temperature distribution. An inner heliosheath proton
distribution function was derived that is 1) consistent with V2 solar
wind plasma observations, and 2) is similar to a $\kappa$-distribution
with index 1.63. The composite inner heliosheath proton distribution
function is a superposition of cold transmitted solar wind protons, a
hot transmitted PUI population, and a very hot PUI population that was
reflected by the cross-shock electrostatic potential at least once
before being transmitted downstream. The composite spectrum possesses
more structure than the $\kappa$-distribution but both distributions
have approximately the same number of protons in the wings of the
distribution (and therefore many more than a corresponding Maxwellian
distribution). Finally, ENA spectra from various directions at 1 AU
generated by either the composite (TS) heliosheath proton distribution
or the $\kappa$-distribution are very similar in intensity, although
some structure is present in the composite case. The spectral shape is
a consequence of the contribution to the ENA flux by primarily
heliosheath transmitted and reflected PUIs. The ENA spectrum is
dominated by transmitted PUI created ENAs in the energy range below 2
keV and reflected PUI created ENAs in the range above 2 keV. This may
give us an opportunity to use IBEX data to directly probe the
microphysics of the TS. The skymaps are dominated by ENAs created by
either transmitted PUIs or reflected PUIs, depending on the energy
range.

IBEX, in completing its first full scan of the sky, created maps of
energetic neutral atom (ENA) flux for energies between 100 eV and 6
keV (McComas et al. 2009a; Schwadron et al. 2009; Funsten et
al. 2009b; Fuselier et al. 2009). The overall flux intensities appear
to be generally within about a factor of two or three of those
predicted by global models of the interaction between the solar wind
(SW) and local interstellar medium (LISM). A most unexpected feature
was the presence in the IBEX ENA maps of a ``ribbon'' that encircles
the sky, passing closer to the heliospheric nose direction in the
south and west than in the north and east. The ribbon represents a
nearly threefold enhancement in ENA flux compared to adjacent parts of
the sky, but the shape and magnitude of the energy spectrum is
primarily ordered by ecliptic latitude rather than its location inside
or outside of the ribbon (Funsten et al. 2009b). This suggests that
ENAs inside the ribbon come from the same population of parent ions.
3D global heliospheric models make it possible to simulate the flux of
ENAs at 1 AU (Fahr \& Lay 2000; Gruntman et al. 2001; Heerikhuisen et
al. 2007; Sternal et al. 2008; Prested et al.  2008; Heerikhuisen et
al. 2008; Izmodenov et al. 2009). The assumptions made by global
models have been refined as new observational data emerged. For
example, the termination shock (TS) crossing by the Voyager 1 \& 2
spacecraft, in 2004 and 2007 respectively (Stone et al. 2005, 2008),
suggested a north-south asymmetry of the heliosphere. The inferred
asymmetry led to new global models with larger than previously thought
interstellar magnetic field (ISMF) strengths (Pogorelov et al. 2007,
2009; Izmodenov et al. 2009). Measurements of Lyman-alpha
back-scattered photons in the nearby SW (Lallement et al. 2005),
suggest asymmetries in the outer heliosheath (OHS) that can be linked
to the plane of the LISM magnetic and velocity vectors – the so-called
``hydrogen deflection plane''. Models confirmed (Izmodenov et
al. 2005; Pogorelov et al. 2008, 2009) that the deflection of
interstellar hydrogen from helium due to the shape of the OHS does
indeed take place primarily in the hydrogen deflection plane. The IBEX
observations enable the first global validation of these models and
their components and, thus, yield insight into the physical processes
that drive the structure and dynamics of the outer heliosphere. The
fact that the ribbon was not predicted by any models suggests that it
is generated by physical processes that have so far been omitted from
models.

The relationship between the ribbon and the region just outside the
heliopause where the ISMF is perpendicular to radial vectors from the
sun, was discovered by the IBEX team (McComas et al.  2009a; Funsten
et al. 2009b; Schwadron et al. 2009) using model results from
Pogorelov et al.  (2009). Several possible explanations for this
correlation were given in those papers, some of which rely on stresses
created by the ISMF near the heliopause to generate regions of
enhanced density which, combined with a local population of
non-isotropic PUIs, may lead to enhanced ENA emissions. However, plots
of the total pressure (magnetic plus thermal) on the surface of the
simulated heliopause display no banded structures related to magnetic
forces or density enhancements at all. Self-consistently coupled
MHD-neutral solution indicate that enhancements in magnetic pressure
and thermal pressure are somewhat anti-correlated, resulting in a
relatively smooth total pressure profile. The underlying physics for
generating the ribbon discovered in the IBEX data must explain a
number of observed features. Firstly, the ribbon appears to be closely
related to the orientation of the magnetic field just outside the
heliopause, in a way that links enhanced ENA flux to regions where the
outer heliosheath magnetic field BOHS is perpendicular to the
heliocentric radial vector r (see figure 4 in McComas et al. (2009a);
figure 3 in Funsten et al. (2009b); and figure 2 in Schwadron et
al. (2009)) such that ${\bf B}_{\rm OHS} \cdot {\bf r} \sim 0$. Secondly,
it needs to explain why the spectrum of ENAs is very nearly the same
inside and outside the ribbon (figure 2 in McComas et
al. (2009a)). Thirdly, it must be based on physical processes that are
excluded from all previous heliospheric models, thereby explaining why
no ENA ribbon feature has been seen in any models of the SW-LISM
interaction.

\begin{figure}[ht]
\sidecaption
\includegraphics[scale=.63]{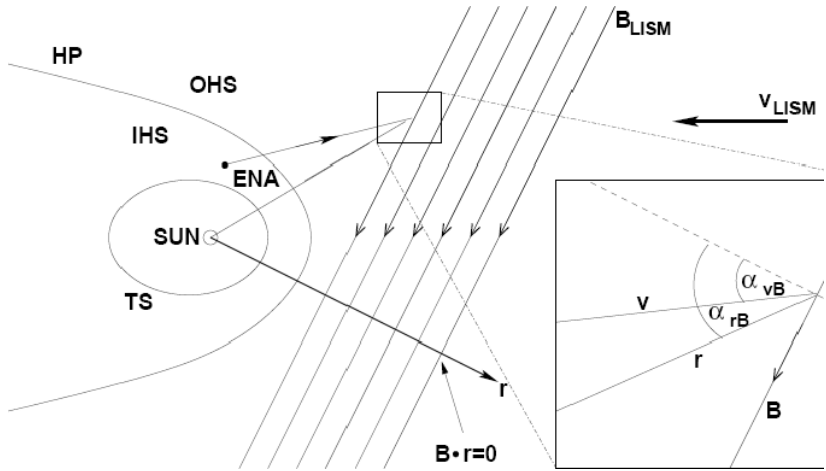}
\caption{ Schematic of the heliosphere in the plane containing the
  ISMF and velocity vectors (BLISM \& VLISM). A primary energetic
  neutral atom (ENA) created in the inner heliosheath (IHS) region
  between the termination shock (TS) and the heliopause (HP) is shown
  as it moves into the outer heliosheath (OHS) whereupon it ionizes
  and becomes an outer heliosheath pickup ion (PUI) that can
  re-neutralize to form a secondary ENA. Note that the OHS magnetic
  field becomes highly warped close to the heliopause (see Pogorelov
  et al., 2009). [Heerikhuisen et al., 2010]}
\label{fig:9} 
\end{figure}

Heerikhuisen et al., 2010 considered the possibility that solar
wind-created neutrals could create pick-up ions (PUIs) in the outer
heliosheath to explain the ribbon of enhanced ENA flux observed by
IBEX. Their approach relies on the fact that the average velocity of
ions in the SW and inner heliosheath (IHS) is anti-sunward, so that
the majority of ENAs propagate away from the sun into the outer
heliosheath. In the region of enhanced interstellar plasma density
surrounding the heliopause, some of these ENAs charge-exchange and
create PUIs in the slow warm subsonic plasma of the outer
heliosheath. These PUIs will initially form a ring-beam distribution,
with a velocity component along the magnetic field. Over time this
distributionwill isotropize by wave-particle interactions (Williams \&
Zank 1994). However, the ring distributed PUIs may charge-exchange
with the fairly dense interstellar hydrogen ($> 0.2 cm^{-3}$),
resulting in a new ``secondary'' ENA. These secondary ENAs have been
included in models before (Izmodenov et al. 2009), but only in an
isotropic way for an axially symmetric heliosphere without an
interstellar magnetic field (ISMF). If ``re-neutralization'' occurs
quickly, the PUI will not have had time to scatter to some random
direction over a complete shell, but rather the secondary ENA will be
directed to some random vector on a partial shell. Furthermore, in
locations where ${\bf B}_{\rm OHS} \cdot {\bf r} \sim 0$, the plane of
the ring about which the shell distribution is forming intersects the
Sun, and leads to an increased ENA flux from these locations (see Fig
9). This mechanism could thus explain the link between the ribbon and
the orientation of the ISMF.

Heerikhuisen et al., 2010 used a 3D steady-state MHD-plasma/kinetic
neutral model of the heliosphere (Heerikhuisen et al. 2008, 2009;
Pogorelov et al. 2008), with uniform SW conditions and a 3 μG ISMF in
the hydrogen deflection plane pointed towards ecliptic coordinate
(224,41).  The LISM boundary conditions are consistent with the
analysis of Slavin \& Frisch (2008). They used a Lorentzian (or
``kappa'') distribution for IHS protons. Plotted in Figure 10 are
all-sky maps of ENA flux for both the simulated and observed data. The
simulated ribbon does not line up exactly, but the offset is almost
certainly due to a slightly ``incorrect'' choice of the ISMF
orientation in the simulation. The observed ribbon, particularly the
southern-most portion, moves slightly at high energies. The
Heerikhuisen et al. simulation reproduces this effect, which can be
attributed to the larger mean free path of high energy primary ENAs
resulting in a ribbon from PUIs at a larger distance into the outer
heliosheath, where the magnetic field orientation is slightly
different.

A second important observation is the absence of a unique spectral
signature associated with the ribbon. The all-sky spectrum predicted
by the Heerikhuisen et al simulations shows that the ribbon appears to
have a locally steeper spectrum, while the observed spectrum shows
almost no change across the ribbon. One reason for a steeper ribbon
spectrum in the simulation is that a ``bump'' is formed at the SW
energy, which is uniform and constant in their simulation while in
reality the bump should be spread over a time-averaged SW energy
profile. This deficiency may be addressed in future models by using a
spectrum that depends on physical processes of energization at the
termination shock as experienced by core SW ions and PUIs (Zank et al.
2010). Using such a composite spectrum would also allow for spectral
indices of less than 1.5 over the IBEX energy range, something that is
not possible with a $\kappa$-distribution (Livadiotis \& McComas
2009).

\begin{figure}[ht]
\sidecaption
\includegraphics[scale=.67]{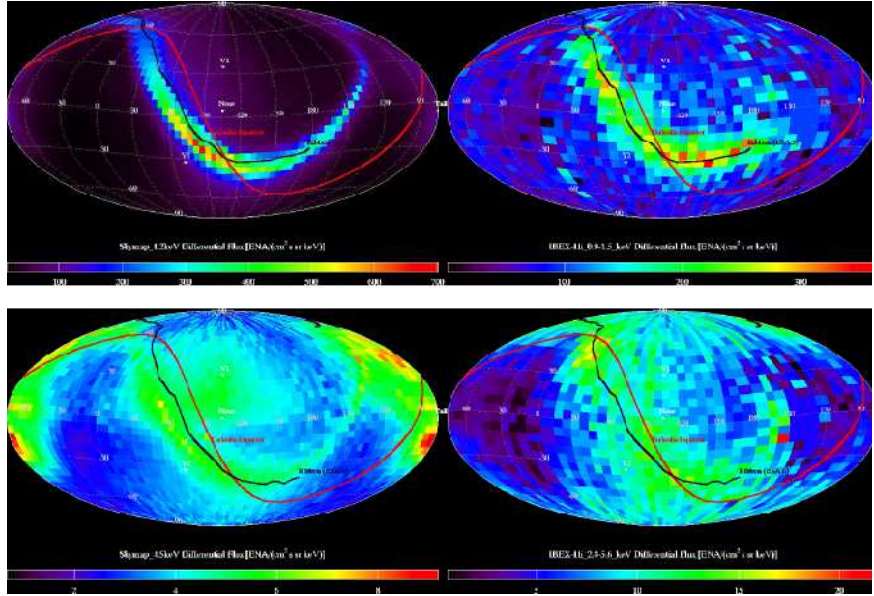}
\caption{ All-sky maps of simulated (left) and observed (right) ENA
  flux at 1.1 keV (top) and 4.5 keV (bottom). The simulation uses a =
  1.63 spectral index for IHS protons and has assumed that all PUIs
  retain partial shell distributions long enough to re-neutralize
  before they isotropize. The red curve is the galactic plane, and a
  best fit to the observed ribbon is shown as a black line. Note that
  the ribbon shifts down slightly at high energies. Units of ENA flux
  are (cm$^2$ s sr keV)$^{-1}$ . [Heerikhuisen et al., 2010]}
\label{fig:10} 
\end{figure}

Finally, A careful comparison with the observed ribbon suggests that
if the Heerikhuisen mechanism is correct, then the ISMF is directed
close to the ecliptic coordinates (224,41) used in their model, and
close to the value (221,39) corresponding to the center of the ribbon
observed in Funsten et al. (2009b).

\section{Conclusions}
We have considered several illustrative examples of the complicated
interplay and coupling between large and small space-time scales, and
slow and fast processes in space plasmas of magnetospheric,
heliospheric and interstellar origin. Laminar and turbulent processes
coexist, and energy transfer manifests itself from small to large and
large to small scales - essentially direct and inverse cascades –
ensuring that a full understanding of complex space plasma systems
requires the proper coupling of disparate scales.

\begin{acknowledgement}
Acknowledgements This study was fulfilled as a part of the Programs of
the Russian Academy of Sciences: ``Origin and Evolution of Stars and
Galaxies'' (P-04), ``Solar Activity and Space Weather'' (P-16, Part 3)
and “Plasma Processes in the Solar System (OFN-15), MSU
Interdisciplinary Scientific Project and supported by the RFBR grants
07-02-00147, NSh-1255.2008.2. DS and GPZ acknowledge the partial
support of NASA grants NNX09AB40G, NNX07AH18G, NNG05EC85C, NNX09AG63G,
NNX08AJ21G, NNX09AB24G, NNX09AG29G, and NNX09AG62G. The authors are
grateful to participants of the IAGA meeting in Sopron who provided
their published materials used for the preparation of this paper.
\end{acknowledgement}


%
%
%

\end{document}